\documentclass[preprint,12pt]{elsarticle}
\usepackage{amssymb}
\usepackage{amsmath}

\journal{Communications in Nonlinear Science and Numerical Simulation}

\begin{document}

\begin{frontmatter}

\title{Coherence properties of cycling chaos}
\author{T. A. Levanova} 
\author{G. V. Osipov}
\address{Department of Control Theory, Nizhni Novgorod State University, 
  Gagarin Av. 23, 606950, Nizhni Novgorod, Russia}

\author{A. Pikovsky}

\address{Institute for Physics and Astronomy, University of Potsdam,
Karl-Liebknecht-Str. 24-25, 14476 Potsdam, Germany}

\begin{abstract}
Cycling chaos is a heteroclinic connection between several chaotic attractors, at which
switching between the chaotic sets occur at growing time intervals. Here we characterize
the coherence properties of these switchings, considering nearly periodic regimes that
appear close to the cycling chaos due to imperfections or to instability. Using numerical
simulations of coupled Lorenz, Roessler, 
and logistic map models, we show that the coherence is high in the case of imperfection 
(so that asymptotically the cycling chaos is very regular), while it is low close to instability
of the cycling chaos.
\end{abstract}

\end{frontmatter}

\section{Introduction}
Heteroclinic cycles have attracted a lot of interest recently. This phenomenon
was first introduced by Guckenheimer and Holmes
\cite{Guckenheimer-Holmes-88,Armbruster88}.
Heteroclinic cycles are stable regimes of switches between meta-stable states,
where these states are observed during longer and longer periods of time, so
that the cycle period grows indefinitely. In the phase space of the system, 
a trajectory passes sequentially through vicinities of saddles, approaching
the limiting closed orbit composed of heteroclinic
pieces~\cite{Ashwin-Chossat-98,Ashwin-Field-99}. 
Physically, heteroclinic cycle appears as a sequential excitation of system's
elements
\cite{afraimovich:1123}. The cases of heteroclinic cycles based on saddle
equilibria
\cite{Krupa-97} or saddle cycles
\cite{Komarov_etal-09,Mikhailov_etal-13,Komarov_etal-13} and heteroclinic orbits
connecting them are well studied.

An interesting variant of a heteroclinic cycle is cycling chaos, first described
by 
Delltitz et al.~\cite{Dellnitz_etal-95} and then studied in 
Refs.~\cite{Palacios-02,Ashwin-97,Ashwin-Field-99}.
Here the saddle states, which a trajectory approaches in the 
course of evolution, are chaotic states. In the simplest
setup one observes sequential periods of chaotic activity of the participating
systems, interrupted 
by epochs of quiescence. While the internal oscillations are strongly irregular,
the switching has a large degree of regularity. 

In this paper we study coherence properties of the cycling chaos.
We characterize the irregularity of the switchings by the variation of their periods,
which is equivalent to the phase diffusion
constant, and study how it depends on the parameters. The paper is organized as
follows. In Section \ref{sec:mcc} we introduce three models of cycling chaos. Two of them
are based
on coupled continuous Lorenz and Rossler oscillators. The third model is a
discrete and based on coupled logistic maps. Then we study the statistical
properties of 
these models in Section~\ref{sec:spc}.

\section{Models of cycling chaos}
\label{sec:mcc}

All models of cycling chaos that we use below are based on standard models of chaos,
interaction of which is organized to ensure cycling. The first model is three
coupled Lorenz systems
\begin{equation}
\begin{aligned}
\dot x_i&=\sigma(y_i-x_i)\\
\dot y_i&=R_ix_i-x_iz_i+\delta\\
\dot z_i&=-bz_i+x_iy_i+\delta
\end{aligned}
\label{eq:lor}
\end{equation}
where $i=1,2,3$ and parameters $\sigma=10$, $b=8/3$ are the standard ones. The
coupling is via the
dependence of parameters $R_i$ on the states of the other oscillators:
\begin{equation}
R_i=28(1-0.08 D z_{i+1}-0.16 z_{i-1})
\label{eq:couplor}
\end{equation}
where the periodicity in the index is assumed (so that $z_4\equiv z_1$,
$z_0\equiv z_3$).
Parameter $D$, as we now show, is responsible for the transition to cycling. The
small parameter $\delta$ describes  non-perfectness of the heteroclinic cycle. 

\begin{figure}
\includegraphics[width=\columnwidth]{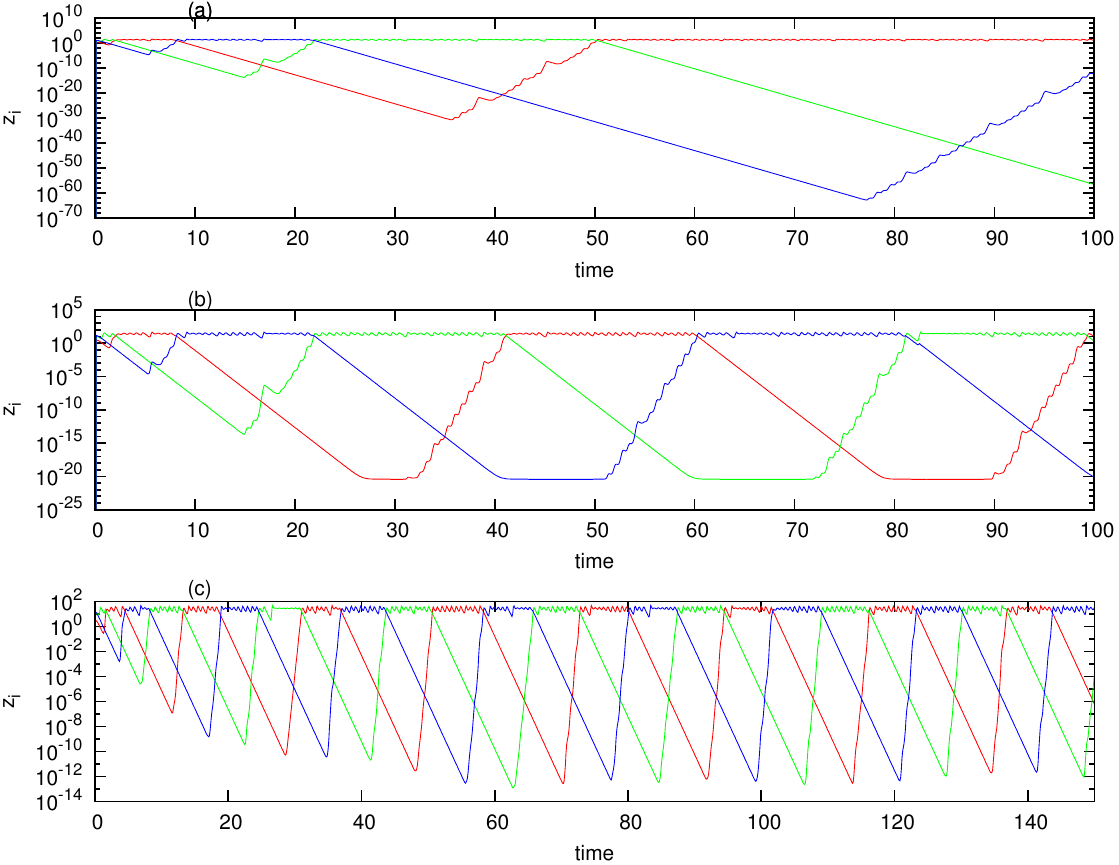}
\caption{Dynamics of Lorenz model~\eqref{eq:lor}. Lines of different colors show
time tependence of variables $z_i$. (a) $D=0.4,\delta=0$: cycling chaos. (b)
$D=0.4,\delta=10^{-20}$: imperfect cycling chaos. (c) $D=0.3,\delta=0$:
heteroclinic contour is unstable
and a cycle in its vicinity is observed. }
\label{fig:lor1}
\end{figure}

The dynamics of system \eqref{eq:lor} is illustrated in Fig.~\ref{fig:lor1}. The
panel (a) shows
the dynamics of variables $z_i$ for $\delta=0$ and $D=0.4$. For these parameters
there exist three invariant chaotic sets, where one Lorenz system is active
(e.g., $x_1,y_1,z_1\neq 0$), while two other vanish (e.g., $x_i=y_i=z_i=0$ for
$i=2,3$). These sets are transversally saddles, forming an attracting
heteroclinic cycle. One can see how a trajectory approaches this cycle, staying
at each of
chaotic sets for longer and longer times. The panel (b) shows the regime for
$D=0.4$ and $\delta=10^{-20}$. Now the non-active Lorenz systems do not vanish,
but the variables have values $\approx \delta$. Correspondingly, the trajectory
approaches a cycling orbit, period of which is bounded from above, tending to
infinity as non-perfectness $\delta$ decreases.  
Finally, in the  panel (c) we show the same system for $\delta=0$ and $D=0.3$.
At this parameter
of coupling the invariant chaotic sets, although present, do not constitute a
stable heteroclinic cycle, but because the repelling normal Lyapunov exponent is
larger than the attracting one, a regime
with a bounded period appears (see \cite{Ashwin-97} for details).

Our goal is to characterize cycles at middle and bottom panels of
Fig.~\ref{fig:lor1} statistically. Together with the Lorenz model
\eqref{eq:lor}, we consider two other models with similar behavior.
A. Palacios~\cite{Palacios-02} has introduced the model of coupled logistic maps
(which we generalize
by introducing the parameter $\delta$ having the same meaning as above)
\begin{equation}
\begin{aligned}
x_{n+1}&=\lambda x_n(1-x_n)+\alpha x_n(y_n)^{0.1}+\delta\\
y_{n+1}&=\lambda y_n(1-y_n)+\alpha y_n(z_n)^{0.1}+\delta\\
z_{n+1}&=\lambda z_n(1-z_n)+\alpha z_n(x_n)^{0.1}+\delta
\end{aligned}
\label{eq:log}
\end{equation} 
Here we fix, following \cite{Palacios-02}, $\lambda=3.8$ and consider $\delta$ as
a parameter of imprefection, and $\alpha$ as a parameter responsible for the
lost of stability of the heteroclinic cycle.

Furthermore, we consider three coupled (similar to the coupling in
\eqref{eq:lor}) Roessler systems
\begin{equation}
\begin{aligned}
\dot x_i&=-y_i-z_i+\delta\\
\dot y_i&=a_iy_i+x_i+\delta\\
\dot z_i&=z_i(x_i-10)+0.02*x_i
\end{aligned}
\label{eq:roes}
\end{equation}
with $a_i=0.15(1-0.006 D(x^2_{i+1}+y^2_{i+1})-0.012 (x^2_{i-1}+y^2_{i-1}))$.
Parameters $\delta$ and $D$ have the same meaning as for the Lorenz
systems~\eqref{eq:lor}.

\section{Statistical properties of cycles}
\label{sec:spc}

Our main aim is to describe the nearly-heteroclinic cycling chaos like in
Fig.~\ref{fig:lor1} statistically. One can see from this figure, that the main
macroscopic irregularity is in different lengths of epochs where one of three
participating subsystems dominate. To characterize this 
we performed long runs to get the local periods of cycles $T_k$,
$k=1,\ldots,10^6$, and from these data we calculated the mean period $\langle
T\rangle$ and the standard deviation $\Delta T=\langle
(T-\langle T\rangle)^2\rangle^{1/2}$. The ratio $\gamma=\frac{\Delta T}{\langle
T\rangle}$ is a dimensionless quantity
characterizing the coherence of the cycle, it is proportional to the phase
diffusion constant normalized by the period.

\begin{figure}[!htb]
\tiny{(a)}\includegraphics[width=0.3\columnwidth]{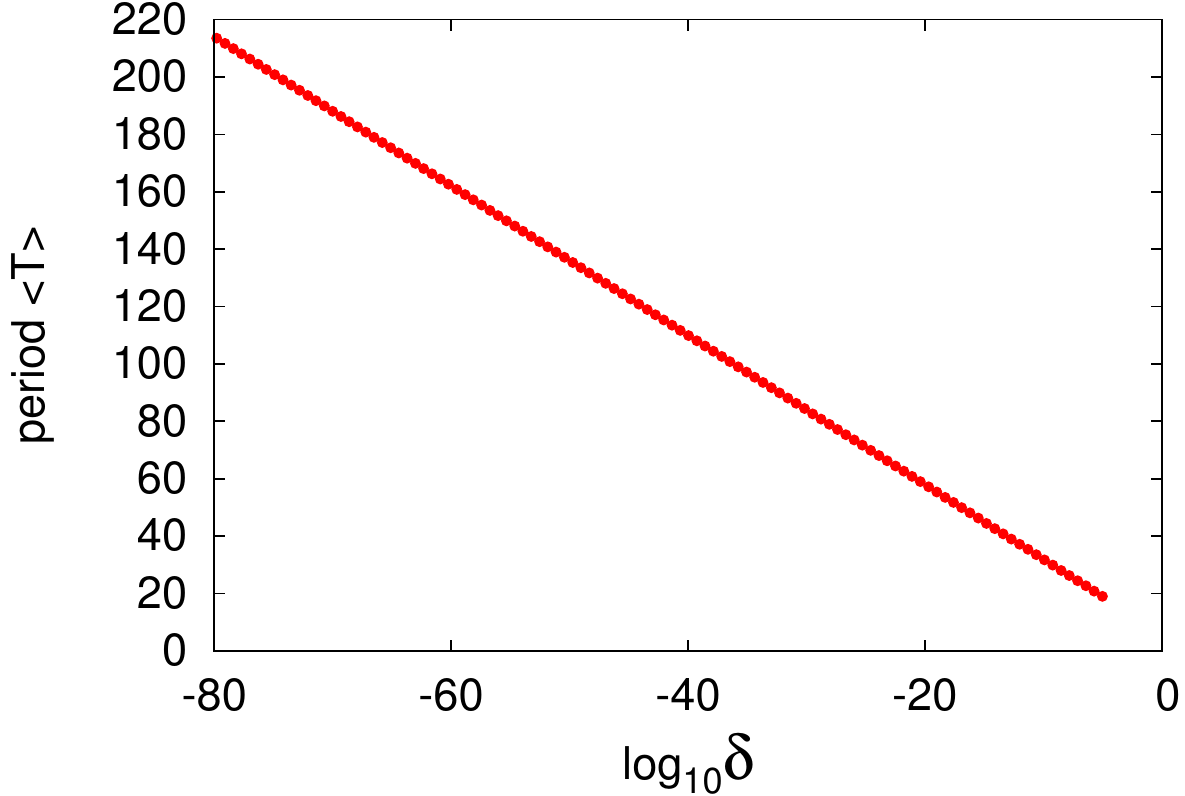}\hfill
\tiny{(b)}\includegraphics[width=0.3\columnwidth]{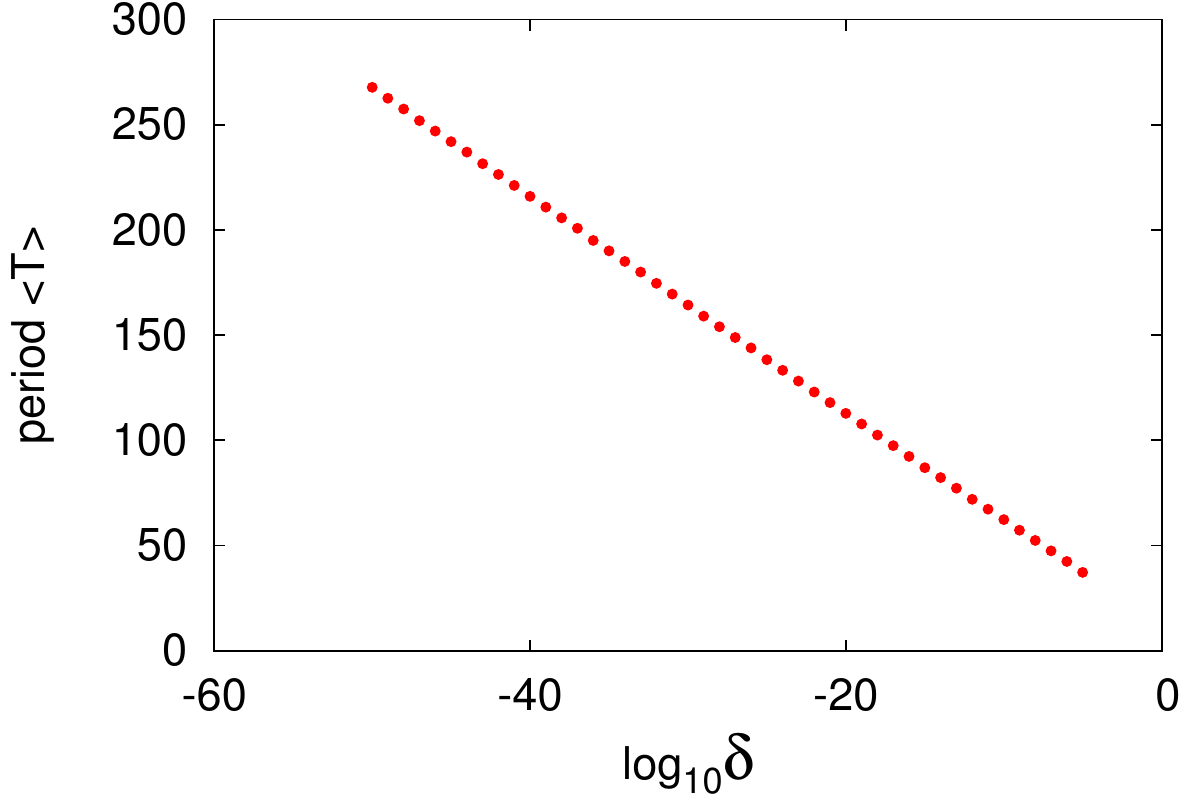}\hfill
\tiny{(c)}\includegraphics[width=0.3\columnwidth]{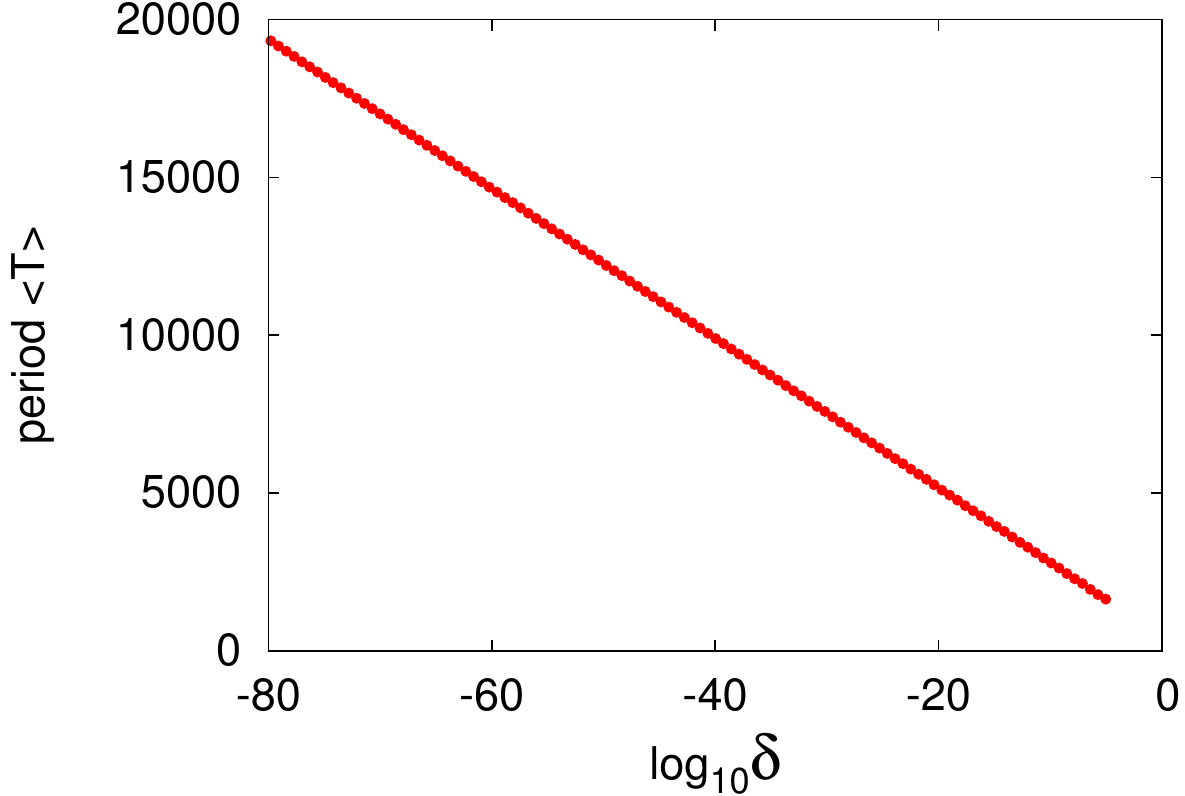}
\caption{Dependencies of the average period on the parameter $\delta$:
(a) Lorenz model~\eqref{eq:lor} with $D=0.4$, (b) Logistic map model~\eqref{eq:log} 
with $\alpha=-3.72$, 
(c) Roessler model~\eqref{eq:roes} with $D=0.8$}
\label{fig:delta1}
\end{figure}

First we present the results for the cycles appearing due to the imperfectness
parameter $\delta$.
In Fig.~\ref{fig:delta1} we show the dependence of $\langle T\rangle$ on
$\log\delta$ for the models considered. As one can expect, the dependence is
linear.

\begin{figure}[!htb]
\tiny{(a)}\includegraphics[width=0.3\columnwidth]{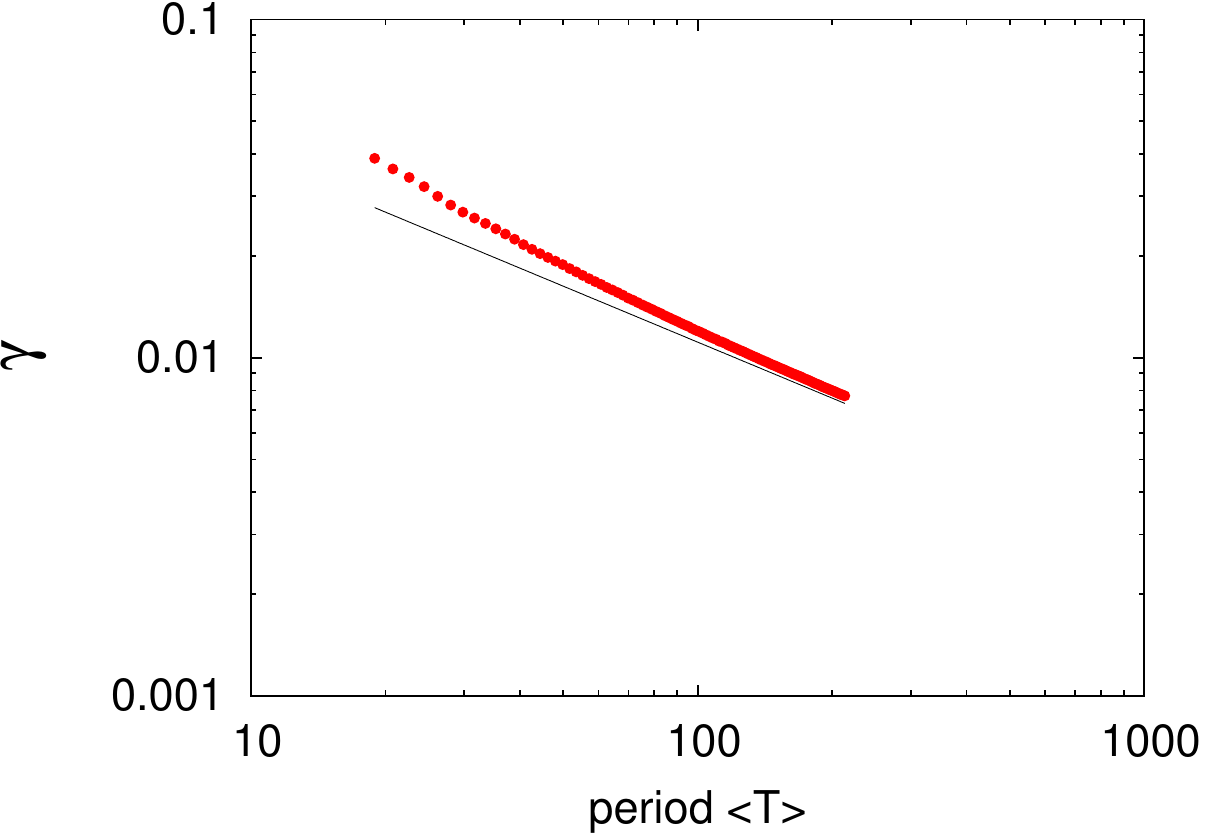}\hfill
\tiny{(b)}\includegraphics[width=0.3\columnwidth]{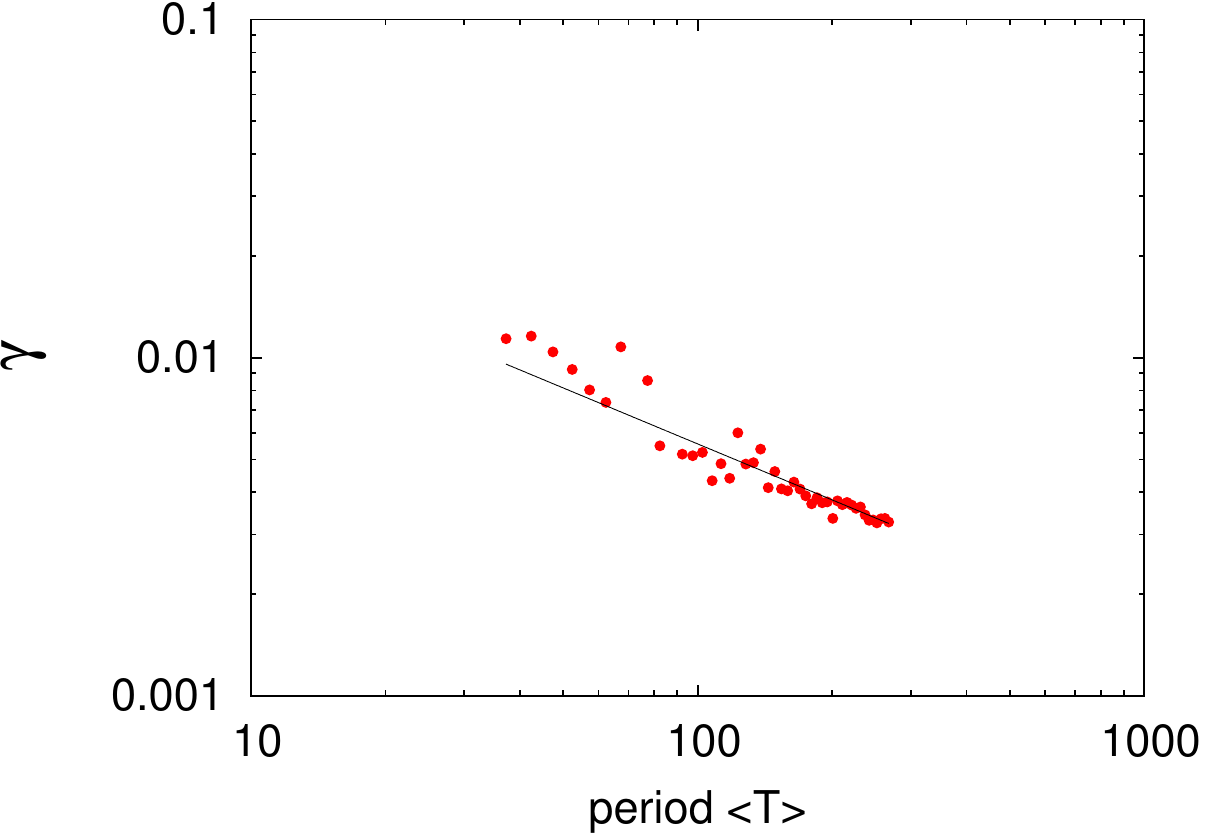}\hfill
\tiny{(c)}\includegraphics[width=0.3\columnwidth]{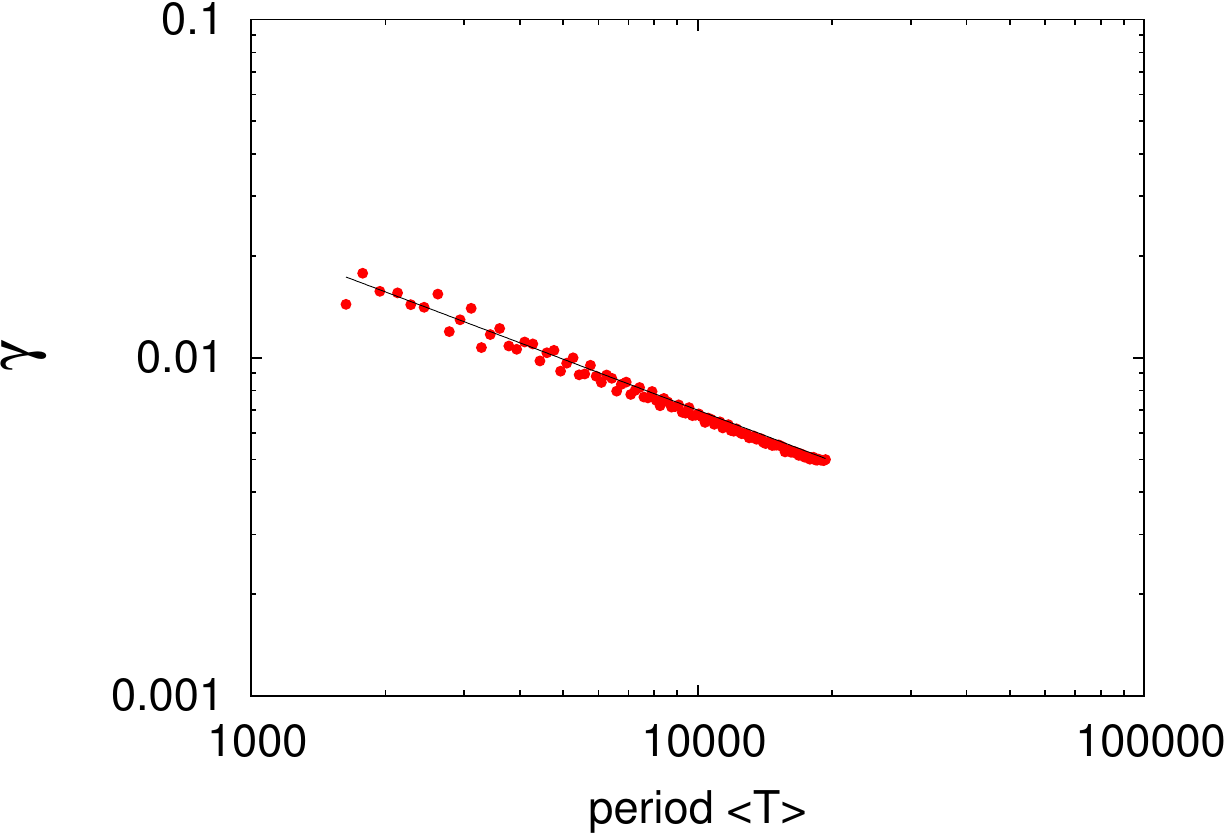}
\caption{Dependencies of the cycle coherence $\gamma$ on the average
period for the same parameters as in Fig.~\ref{fig:delta1}:
(a) Lorenz model~\eqref{eq:lor}, (b) Logistic map model~\eqref{eq:log}, (c)
Roessler model~\eqref{eq:roes}. The slopes of the lines in (a),(b) are 0.55, in
(c) the slope is 0.5.}
\label{fig:delta2}
\end{figure}

The coherence parameter $\gamma$ is presented in
Fig.~\ref{fig:delta2}. One can see
that in all cases the coherence improves as the period of the cycle increases,
roughly $\gamma=\frac{\Delta T}{\langle T\rangle}\sim \langle T\rangle^{-1/2}$. 
This indicates that asymptotically
for long cycles the coherence is large:
chaotic epochs do not contribute significantly to the phase diffusion.

\begin{figure}
\tiny{(a)}\includegraphics[width=0.3\columnwidth]{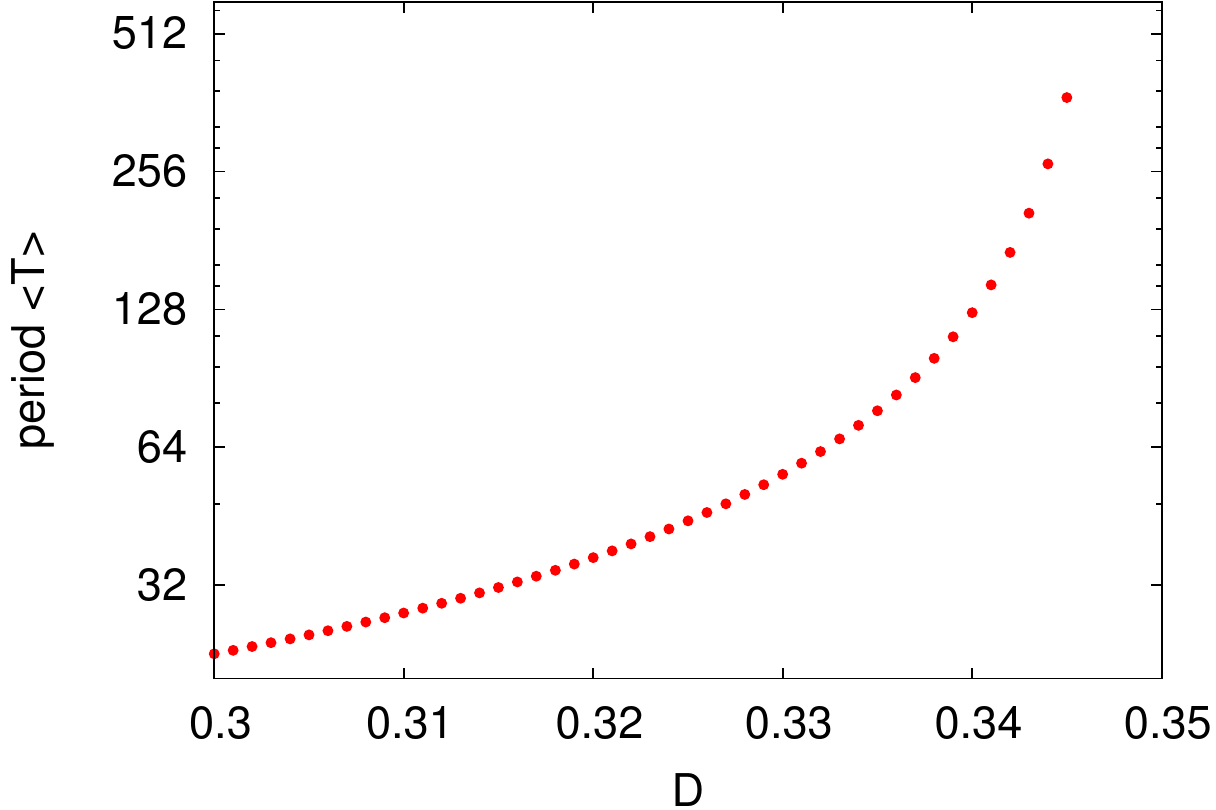}\hfill
\tiny{(b)}\includegraphics[width=0.3\columnwidth]{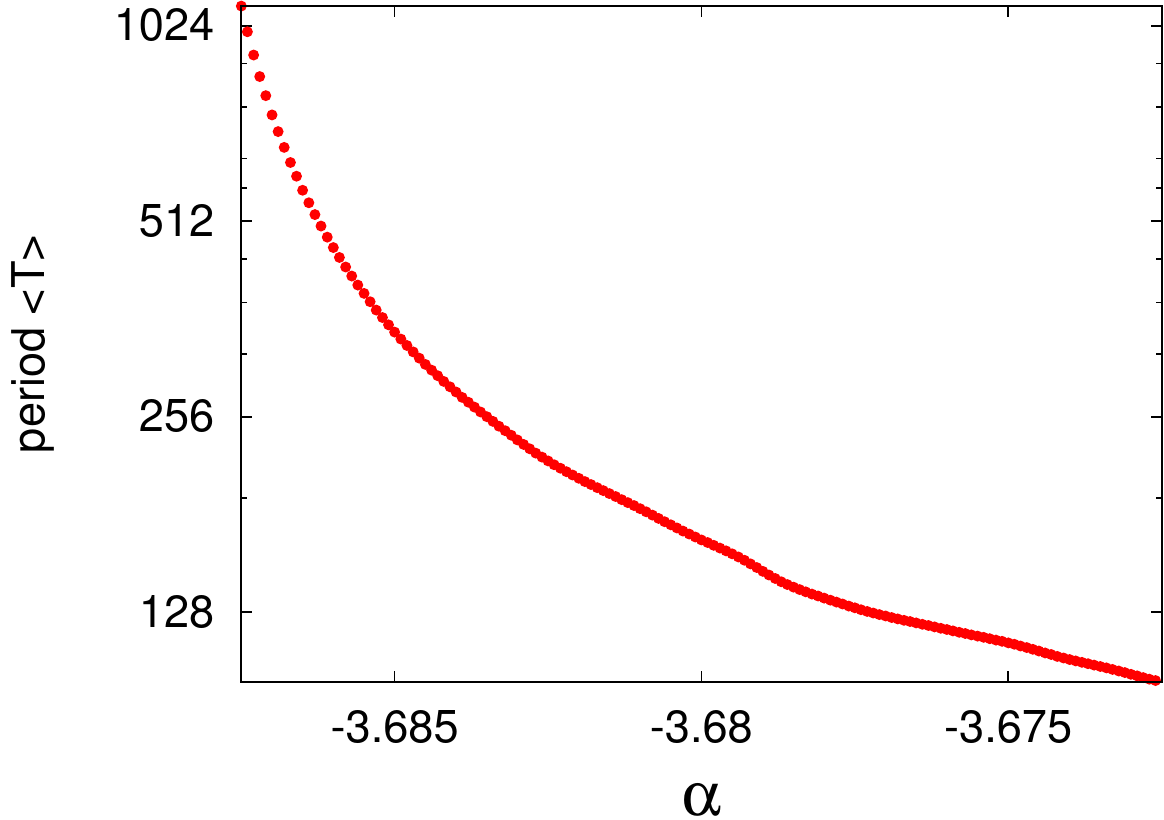}\hfill
\tiny{(c)}\includegraphics[width=0.3\columnwidth]{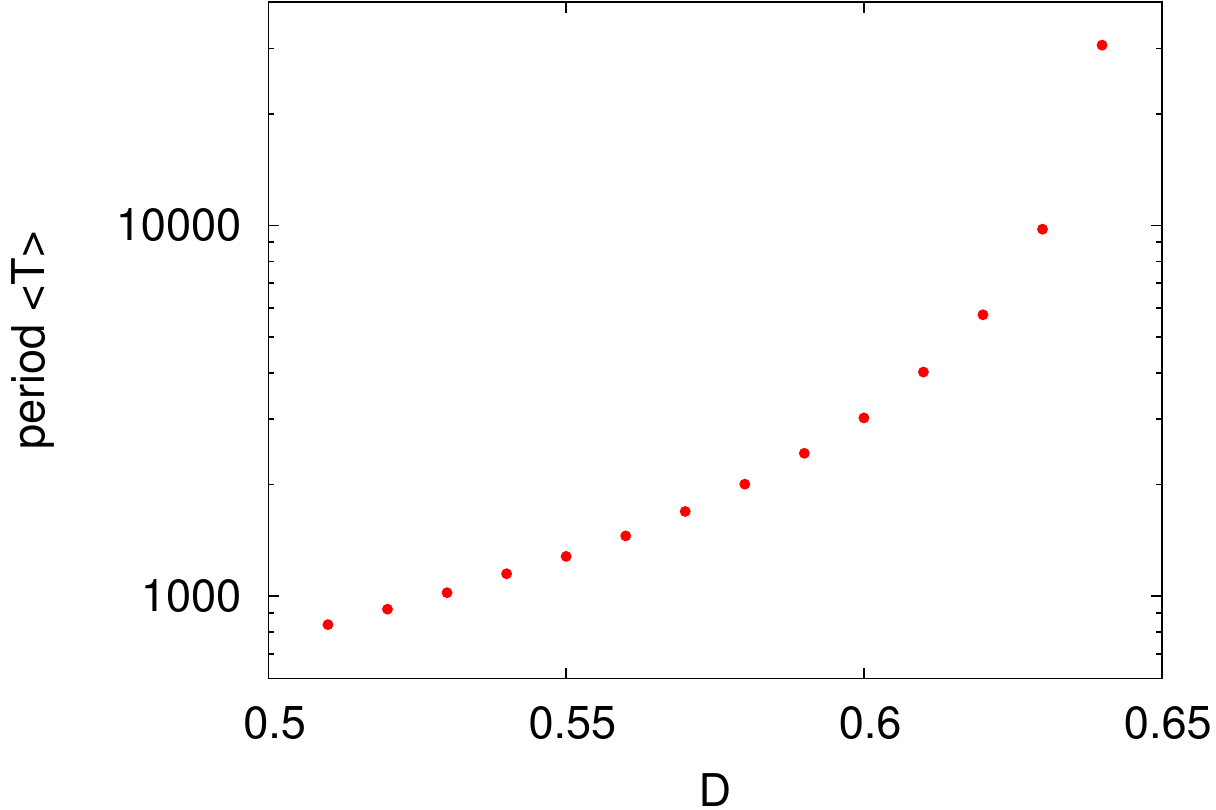}
\caption{Variation of the chaotic cycle period near the transition to a 
chaotic homoclinic cycle, for $\delta=0$. (a): Lorenz model, dpendence on parameter $D$. (b):
Logistic map
model, dependence on parameter $\alpha$. (c): Roessler model, 
dependence on parameter $D$.}
\label{fig:d1}
\end{figure}

\begin{figure}
\tiny{(a)}\includegraphics[width=0.3\columnwidth]{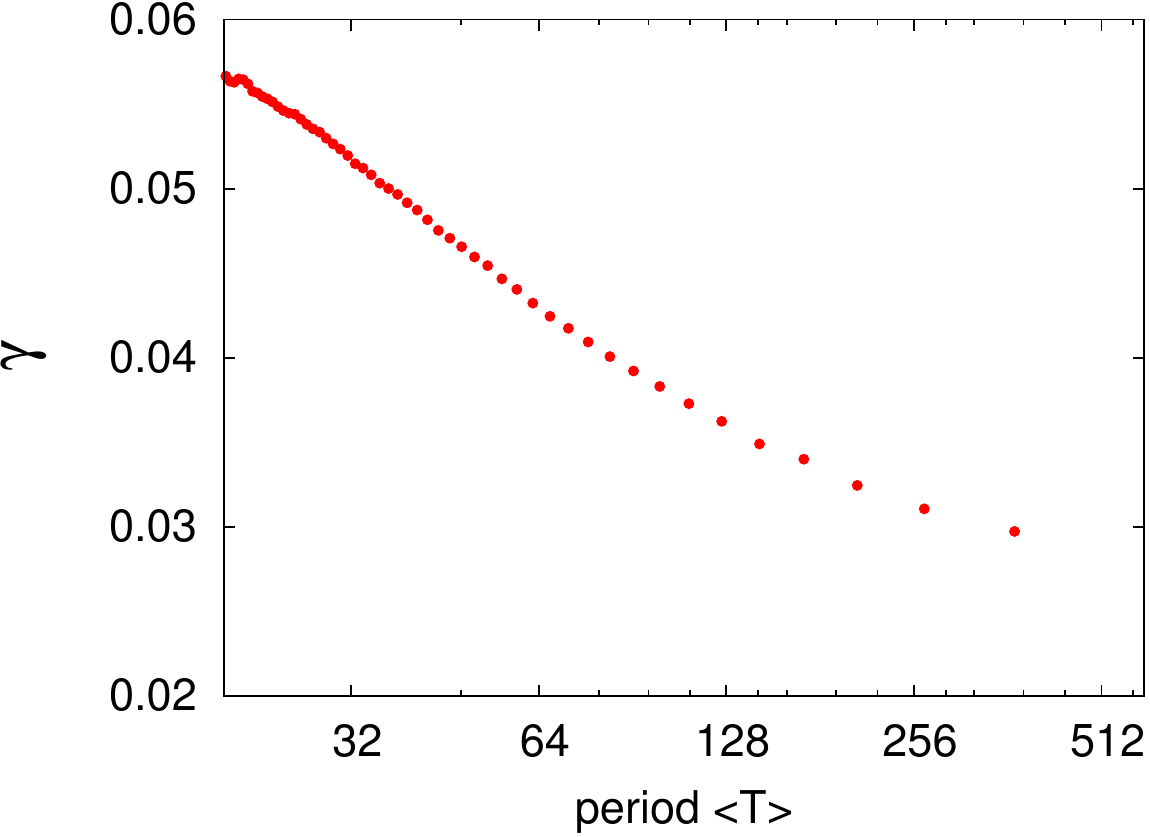}\hfill
\tiny{(b)}\includegraphics[width=0.3\columnwidth]{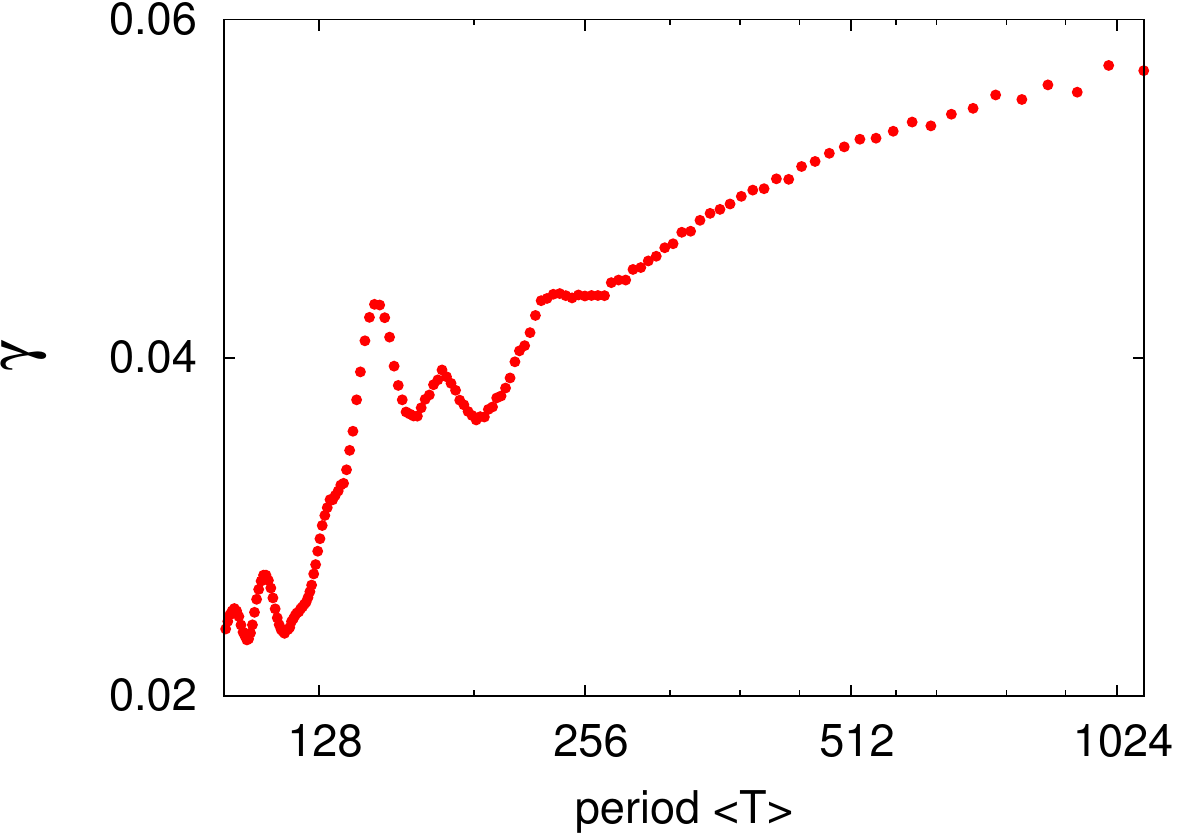}\hfill
\tiny{(c)}\includegraphics[width=0.3\columnwidth]{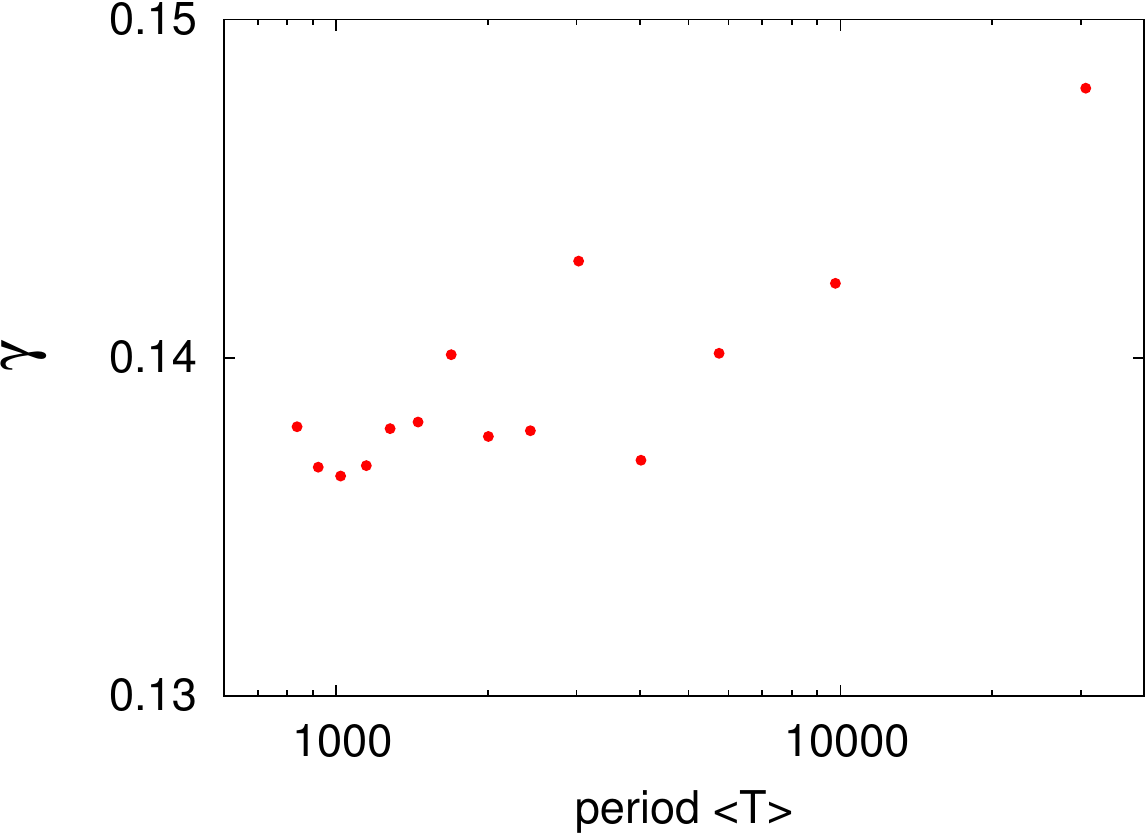}
\caption{Dependencies of the cycle coherence $\gamma$ on the average
period
for the data presented in Fig.~\ref{fig:d1}:
(a) Lorenz model, (b) Logistic map model, (c) Roessler model. Notice that
contrary to 
the Fig.~\ref{fig:delta2}, here on all panels the vertical axis is linear and
not 
logarithmic.}
\label{fig:d2}
\end{figure}

In Figs.~\ref{fig:d1},\ref{fig:d2} we report similar calculations for the second
type of nearly-heteroclinic cycle, presented in Fig.~\ref{fig:lor1}(c).
Figure~\ref{fig:d1}  shows the dependencies
of the average period on the coupling parameters ($D$ for Lorenz and Roessler
models, $\alpha$ for the logistic maps).  Near the criticality the period grows,
although the precise law of this growth is not easy to follow, because the
critical values of coupling are known only approximately. The coherence
parameter
is presented in Fig.~\ref{fig:d2}, again as a function of the period. One can
see that the behavior for the three models is different: for the Lorenz model
the coherence improves with the period, while for the logistic maap model it
decreases; for the Roessler model no significant changes are observed. In all
three cases the coherence parameter $\gamma=\frac{\Delta T}{\langle T\rangle}$ does not
appear to follow a power-law, rather it seems that it remains bounded as
$\langle T\rangle\to\infty$. (For the only candidate for a power-law, the Lorenz system,
we plotted the data of Fig.~\ref{fig:d2} in a log-log representation and observed
that it does not follow a power-law, but the slope decreases 
from $\approx -0.3$ to $\approx -0.15$, with a tendency to saturation of the coherence.)
The results of the simulations for all three models show that the coherence of the long cycles
close to the transition is rather weak.

The difference in the behavior of the coherence parameter can be explained as
follows.
In the case of the imperfect heteroclinic cycle ($\delta\neq 0$,
Figs.~\ref{fig:delta1},\ref{fig:delta2}) the minimal level of activity is fixed
by the value of $\delta$, the period of 
the cycle is basically determined by the time needed to reach the activity
starting from $\delta$,
and the variations of this time are relatively small. In the case where the
finite cycle  appears 
from the stable heteroclinic one at the critical coupling, where the negative
transverse Lyapunov 
exponent becomes equal to the positive one, the minimal level of activity is
determined by the 
difference of these exponents. Because the Lyapunov exponents in chaos
fluctuate, this minimal level
experiences significant fluctuations as well. This can be clearly seen as
fluctuations of minima in 
Fig.~\ref{fig:lor1}(c). Thus, the time needed to reach activity from the minimal
level fluctuates as well, which results in relatively weak coherence.

\section{Conclusion}

Cycling chaos is an interesting example of nontrivial dynamics ``on top'' of chaos.
In other examples, such as chaos-induced 
diffusion~\cite{Geisel-Nierwetberg-82,Just_et_al-01}, chaotic motion enters 
explicitely. Here in the ideal case chaos lives on the invariant manifolds of three participating 
oscillators, and in perturbed case only finite epochs 
of chaotic dynamics are observed. This
makes analytical calculations of statistical properties of cycling chaos 
a challenging problem for the future, 
therefore in this communication we followed a numerical approach. We simulated three
models, two are coupled continuous-time chaotic systems (Lorenz and Roessler models), and
one is coupled logistic maps. Our approach was not to characterize cycling chaos, as it 
is a nonstationary process, but to characterize the coherence of the nearly periodic oscillations
that appear if the cycling chaos is disturbed, and to check the properties of the coherence as the
cycle approaches pure heteroclinic
cycling chaos, and its average period  grows.
In all cases we observed a clear difference 
between the situations when the cycling chaos was destroyed by an imperfection in the equations (the chaotic 
manifolds are no more invariant ones), and when it becomes unstable as the 
instability of the manifolds becomes stronger than the attraction to them. In the former case the coherence
is high and the deviations from the average period follow the power law, growing roughly as a 
square root of the period. This means that the relative coherence increases with the period of the cycle. 
When the cycling chaos loses its stability, we observe that the deviations are roughly proportional 
to the period, what means that the relative coherence remains finite. The properties of cycle coherence are 
important, e.g., for synchronizability of the cycling chaos, what is the subject of future work.

\bigskip
\textbf{Acknowledgements} 
We would like to thank  Russian Federal Program “Scientific and Scientific-Educational Brainpower of Innovative Russia” for 
2009--2013 (contract No.14.B37.21.0863). 
A.P. thanks
 Nizhni Novgorod State University for warm hospitality.


\begin{thebibliography}{10}
\expandafter\ifx\csname url\endcsname\relax
  \def\url#1{\texttt{#1}}\fi
\expandafter\ifx\csname urlprefix\endcsname\relax\def\urlprefix{URL }\fi
\expandafter\ifx\csname href\endcsname\relax
  \def\href#1#2{#2} \def\path#1{#1}\fi

\bibitem{Guckenheimer-Holmes-88}
J.~Guckenheimer, P.~Holmes, {Structurally stable heteroclinic cycles}, Math.
  Proc. Camb. Phil. Soc. 103 (1988) 189--192.

\bibitem{Armbruster88}
D.~Armbruster, J.~Guckenheimer, P.~J. Holmes, Heteroclinic cycles and modulated
  travelling waves in systems with {O2} symmetry., Physica D 29 (1988)
  257--282.

\bibitem{Ashwin-Chossat-98}
P.~Ashwin, P.~Chossat, Attractors for robust heteroclinic cycles with continua
  of connections, J. Nonlin. Sci. 8 (1998) 103--129.

\bibitem{Ashwin-Field-99}
P.~Ashwin, M.~Field, Heteroclinic networks in coupled cell systems, Arch.
  Rational Mech. Anal. 148 (1999) 107--143.

\bibitem{afraimovich:1123}
V.~S. Afraimovich, V.~P. Zhigulin, M.~I. Rabinovich, On the origin of
  reproducible sequential activity in neural circuits, Chaos 14~(4) (2004)
  1123--1129.

\bibitem{Krupa-97}
M.~Krupa, Robust heteroclinic cycles, J. Nonlinear Sci. 7 (1997) 129--176.

\bibitem{Komarov_etal-09}
M.~A. Komarov, G.~V. Osipov, J.~A.~K. Suykens, equentially activated groups in
  neural networks, EPL 86 (2009) 60006.

\bibitem{Mikhailov_etal-13}
A.~O. Mikhaylov, M.~A. Komarov, T.~A. Levanova, G.~V. Osipov, Sequential
  switching activity in ensembles of inhibitory coupled oscillators, EPL 101
  (2013) 20009.

\bibitem{Komarov_etal-13}
M.~A. Komarov, G.~V. Osipov, C.~S. Zhou, Heteroclinic contours in oscillatory
  ensembles, Phys. Rev. E 87 (2013) 022909.

\bibitem{Dellnitz_etal-95}
M.~Dellnitz, M.~Field, M.~Golubitsky, A.~Hohmann, J.~Ma, Cycling chaos, IEEE
  Trans. Circuits Systems 42~(10) (1995) 3821--823.

\bibitem{Palacios-02}
A.~Palacios, Cycling chaos in one-dimensional coupled iterated maps, Int. J.
  Bif. Chaos 12~(8) (2002) 1859--1868.

\bibitem{Ashwin-97}
P.~Ashwin, Cycles homoclinic to chaotic sets; robustness and resonance, Chaos 7
  (1997) 207--220.

\bibitem{Geisel-Nierwetberg-82}
T.~Geisel, J.~Nierwetberg, {Onset of diffusion and universal scaling in chaotic
  systems}, Phys. Rev. Lett 48~(1) (1982) 7--10.

\bibitem{Just_et_al-01}
W.~Just, H.~Kantz, C.~R\"odenbeck, M.~Helm, Stochastic modelling: {R}eplacing
  fast degrees of freedom by stochastic processes, J. Phys. A: Math., Gen. 34
  (2001) 3199.

\end{thebibliography}

\end{document}